\documentclass[doublecol]{epl2}
\usepackage{graphicx}
\usepackage{bm}
\newcommand{\rmd}{\rm d}
\newcommand{\ie}{{\em i.e. }}

\newcommand{\be}{\begin{equation}}
\newcommand{\ee}{\end{equation}}
\newcommand{\ba}{\begin{array}}
\newcommand{\ea}{\end{array}}

\title{Measuring Nanoscale Stress Intensity Factors with an Atomic Force Microscope}

\author{Kun Han\inst{1} \and Matteo Ciccotti\inst{2}\thanks{E-mail: \email{matteo.ciccotti@univ-montp2.fr}} \and St\'{e}phane Roux\inst{1}\thanks{E-mail: \email{stephane.roux@lmt.ens-cachan.fr}}}
\shortauthor{K. Han \etal}

\institute{
  \inst{1} Laboratoire de M\'{e}canique et Technologie-Cachan,
ENS de Cachan/CNRS-UMR 8535/Universit\'{e} Paris 6/PRES UniverSud Paris,
61 avenue du Pr\'{e}sident Wilson, F-94235 Cachan Cedex, France.\\

  \inst{2} Laboratoire des Collo\"{\i}des, Verres et Nanomat\'{e}riaux,
Universit\'{e} Montpellier 2, CNRS, Montpellier, France
}

\abstract{
Atomic Force Microscope images of a crack intersecting the free
surface of a glass specimen are taken at different stages of
subcritical propagation. From the analysis of image pairs, it is
shown that a novel Integrated Digital Image Correlation technique
allows to measure stress intensity factors in a quantitative fashion.
Image sizes as small as 200 nm can be exploited and the surface displacement fields
do not show significant deviations from linear elastic solutions
down to a 10 nm distance from the crack tip.
Moreover, this analysis gives access to the out-of-plane displacement of the free
surface at the crack tip.}


\begin{document}

\maketitle



The mechanisms of subcritical crack propagation in silicate glasses have been
the subject of extensive research (cf.~\cite{Freiman,Ciccotti} for recent
reviews). The development of advanced AFM techniques to probe {\em in-situ}
crack propagation or {\em post-mortem} crack surface morphologies have recently
led to explore these mechanisms at their relevant nanometric scale leading to
remarkable observations on plastic crack tip damage~\cite{Celarie2004,Fett08},
stress-induced ion migration~\cite{Celarie2007} and capillary condensation
inside the crack tip cavity~\cite{Grimaldi}. However, a proper understanding of
these observations in terms of sound mechanical modeling has been hampered by
the lack of a technique to measure the stress and strain fields at the crack tip
with a nanometric resolution, and the proposed interpretations are still subject
of debate \cite{Ciccotti}. Another weakness of the {\em in-situ} AFM
observations is their limitation to access the external surface of the sample,
that is intersected by the propagating crack front with a small tilt angle, thus
making the local three-axial condition of stress not trivial \cite{Dimitrov}.
The present letter reports a concomitant solution of all these problems. By
extending the Digital Image Correlation technique to AFM topographical images of
{\em in-situ} crack propagation in a silica glass it was possible (1) to
characterise properly the 3D surface displacement fields with nanometric
resolution, (2) to show that they can be fitted by appropriate elastic solution
down to a 10 nm distance from the crack tip and (3) that the locally measured
surface stress-intensity factor is in agreement with the macroscopically
measured value within 5\% uncertainty.

Digital Image Correlation (DIC) is a technique which allows one to measure
displacement fields by matching a reference with a deformed image, most often an
optical image.  Introduced a long time ago in the field of solid mechanics by
Sutton \etal~\cite{Sutton1}, this technique has known a very rapid development
due to the wide availability of digital imaging devices, and the increasing
performance and reliability of analyses.  In particular for cracks, since the
pioneering work of Sutton~\cite{Sutton2}, recent
works~\cite{RH_IJF,Yoneyama,RRH_JPD} have shown that stress intensity factors
can be estimated very accurately either directly through Integrated DIC (IDIC)
which incorporates analytic crack fields~\cite{RH_IJF}, or through a tailored
post-processing of the displacement field~\cite{RH_IJF,Yoneyama,Rethore_JMPS}.
However, most applications of DIC deal with optical images, and very
few with Atomic Force Microscopy (AFM) images. Among those, the most
advanced published works concern thin polycrystalline silicon
films~\cite{Chasiotis1,Cho1,Cho2,Cho3,Cho4}. In particular, the
measurement of a uniform strain field from AFM images at different
scales down to scale $1\times 2$ $\mu$m$^2$ images~\cite{Cho3}
allowed to estimate the elastic properties of polycristalline Silicon
used for MEMS.  More recently, stress intensity factors were
evaluated for cracks propagating in such films~\cite{Cho4}. AFM was
used to locate the crack geometry.
Stress Intensity Factors (SIFs) were then indirectly estimated
through an elastic finite element simulation of the test and specimen
geometry. In a different spirit, a recent study by Xu \etal~\cite{Xu}
investigated systematic bias such as drift and distorsion of large
AFM images ($6.4\times6.4$ $\mu$m$^2$) on unstrained samples in order
to estimate (and reduce) strain measurement uncertainties. These two
sets of studies have demonstrated the feasibility of using DIC on AFM
images, mostly for simple strain fields and large scanned regions.

The present work is aimed at a direct determination of stress
intensity factors through DIC at a much lower scale than previously
considered, (from $1\times1$ $\mu$m$^2$ down to $200\times200$
nm$^2$, \ie about 100 times smaller than in previously reported
studies~\cite{Cho4}), without any recourse to numerical
finite-element simulations. This analysis also provides information
on inelastic processes at play in the crack tip neighborhood through
unresolved discrepancies after accounting for a simple linear
elastic-brittle behavior. Applying DIC requires to overcome a
number of very significant challenges:
As the size of the region of interest is reduced, the noise
level of AFM images becomes very significant.
Moreover, images consist in topographic measurements.  In contrast
    with classical speckle patterns which are simply advected by
    the displacement field, the topography is affected by out-of-plane
    motion of the free surface.  In the present study, we
    will see that the typical peak-to-valley roughness of say 1
    $\mu$m size images is of order 3 nm.  The out-of-plane
    displacement difference due to a crack at the critical stress
    intensity factor will be estimated below as being of order 2
    nm, and thus cannot be ignored.  Thus, the very foundation of
    DIC, \ie the conservation of optical flow, has to be
    revisited and generalized to account for this effect.

\begin{figure}
\centering
\includegraphics[width=0.45 \textwidth]{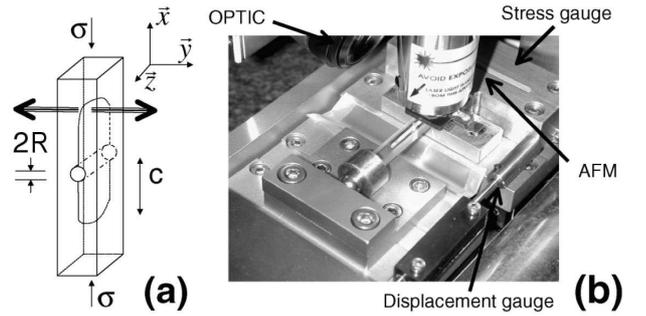}
\caption{Experimental setup: (a) Sketch of the DCDC geometry; (b) picture of the experiment.}
\label{fig:Setup}
\end{figure}

In the present experiments, fractures were initiated and propagated
on a Double Cleavage Drilled Compression (DCDC)
set-up (cf.\ Fig.\ \ref{fig:Setup}) at a constant temperature of $(23 \pm 1)^\circ$C
in a leak-proof chamber under an atmosphere composed of pure nitrogen
and water vapor at relative humidity $(23 \pm 1)$\%.
The DCDC test set-up is particularly
convenient for these studies due to its excellent stability and
several theoretical modeling were devoted to it~\cite{Pallares,Fett09}.
The parallelepipedic DCDC samples ($4\times4\times40$ mm$^3$, $\pm$ 10
$\mu$m) of pure fused silica glass (Suprasil 311, Heraeus,
Germany) were polished to a RMS roughness of 0.5 nm (for an area of
1x1 $\mu$m$^2$) and a hole of radius $R=(500 \pm 10)$ $\mu$m was
drilled at their center to trigger the initiation of the two
symmetric fractures of length $c$. AFM observations
are done in tapping mode on a D3100 from Veeco Metrology Inc., Santa Barbara, CA.
The details of the setup and techniques can
be found in \cite{Celarie2004,Grimaldi}. After an extensive
stabilization consisting of several hours of stationary propagation and imaging,
all the images presented here were acquired within
two-hours, so that crack propagation conditions can be
considered as stable at a propagation velocity $v = (0.7 \pm 0.1)$
nm/s and AFM drifts are minimized.
For an applied force $F = (1844 \pm 4)$ N and an average crack
length $c = (6145 \pm 10)~\mu$m, the SIF can be estimated to $K_I =
(0.39 \pm 0.02)$ MPa.m$^{1/2}$ according to Ref.~\cite{Pallares}.

Let us denote by $f(\bm x)$ and $g(\bm x)$ the topographic images of
the reference and deformed states, where $\bm x$ are the coordinates
in the observation plane. $f$ and $g$ are the height of the observed
surface in nanometers. The generalization of the so-called ``optical
flow conservation'' which relates the two images can be written
  \be
  g(\bm x)=f(\bm x+\bm u(\bm x))+v(\bm x)
  \ee
where $\bm u$ (resp. $v$) denotes the in-plane (resp. out-of-plane)
displacement field.

In order to develop a robust method, we resort to an ``Integrated'' DIC (IDIC)
approach as introduced in Ref.~\cite{RH_IJF}.  The displacement field $(\bm
u(\bm x),v(\bm x))$  is searched for as a combination of a few expected fields
$\Phi_n(\bm x)$
  \be
  (\bm u(\bm x),v(\bm x))=\sum_{i=1}^{10} a_i \bm \Phi_i(\bm x)
  \ee
Those fields consist in rigid body motions $1\le n\le 6$ and mode I crack
displacement fields detailed in the following: The latter fields are well known
in either plane stress or plane strain conditions~\cite{Williams}, which
unfortunately do not strictly apply for a crack front emerging on the free
surface of a 3D sample~\cite{Fett09}. Nevertheless, the in-plane component $\bm
u(\bm x)$ can be well approximated by the plane stress condition. We introduce
the notation $\bm\varphi(\bm x)$ for such an in-plane displacement field for a
crack tip located at the origin, and crack path along the $x<0$ semi-axis. The
displacement field normal to the free surface is not known analytically.  The
latter, for a unit mode I stress intensity factor, is denoted $\psi(\bm x)$ for
the same reference geometry. Because of their different status, the in-plane and
out-of-plane displacements will be treated in the sequel as two independent
fields.

An additional difficulty is that as we follow a crack propagation
under a constant load, the crack tip is present in both images
although at different locations. Let us call $\bm x_0$ and $\bm x_1$
the crack tip positions in $f$ and $g$ respectively.  The
displacement field to be applied to the reference image thus consists
in closing the crack at point $\bm x_0$ and opening it at $\bm x_1$.
Thus the $7^{th}$ and $9^{th}$ basis fields are $\bm\varphi(\bm x-\bm
x_0)$ and $\bm\varphi(\bm x-\bm x_1)$ respectively, while the
$8^{th}$ and $10^{th}$ fields are the out-of-plane components
$\psi(\bm x-\bm x_0)$ and $\psi(\bm x-\bm x_1)$ respectively.  For a
steady state propagation under the same stress intensity factor, we
can write $a_7=-a_9$ and $a_8=-a_{10}$. Those equalities can be
enforced using Lagrange multipliers. Thus the problem consists in
evaluating the above 10 unknowns, $a_n$, (8 degrees of freedom)
through a weak form of the extended optical flow conservation to be
minimized:
  \be
  {\cal T}=\int_{\cal D} \left(g(\bm x)
  -f\left(\bm x+
  a_i\bm \Phi_i^{(t)}(\bm x)\right)-
  a_j\Phi_j^{(n)}(\bm
  x)\right)^2\rmd^2 \bm x
  \ee
where $\bm \Phi_i^{(t)}$ and $\Phi_i^{(n)}$ designate respectively the in-plane
and out-of-plane displacement. When suitably normalized (using the pixel size
and the elastic properties of the silica glass specimen), we see that the SIF is
directly estimated as the $a_7$ parameter value. The problem as it is written is
strongly non-linear because of the occurrence of unknowns as arguments of the
topography $f$. To avoid secondary minima trapping the determination of the
displacement is performed hierarchically from the minimization of the above
$\cal T$ functional from coarse-grained images down to the original images. Each
minimization is performed iteratively resorting to a linearization in the
unknowns $a_n$.  This procedure is similar to the one introduced by Hild
\etal~\cite{Hild02}.

\begin{figure}[!ht]
a)\includegraphics[width=0.45\columnwidth]{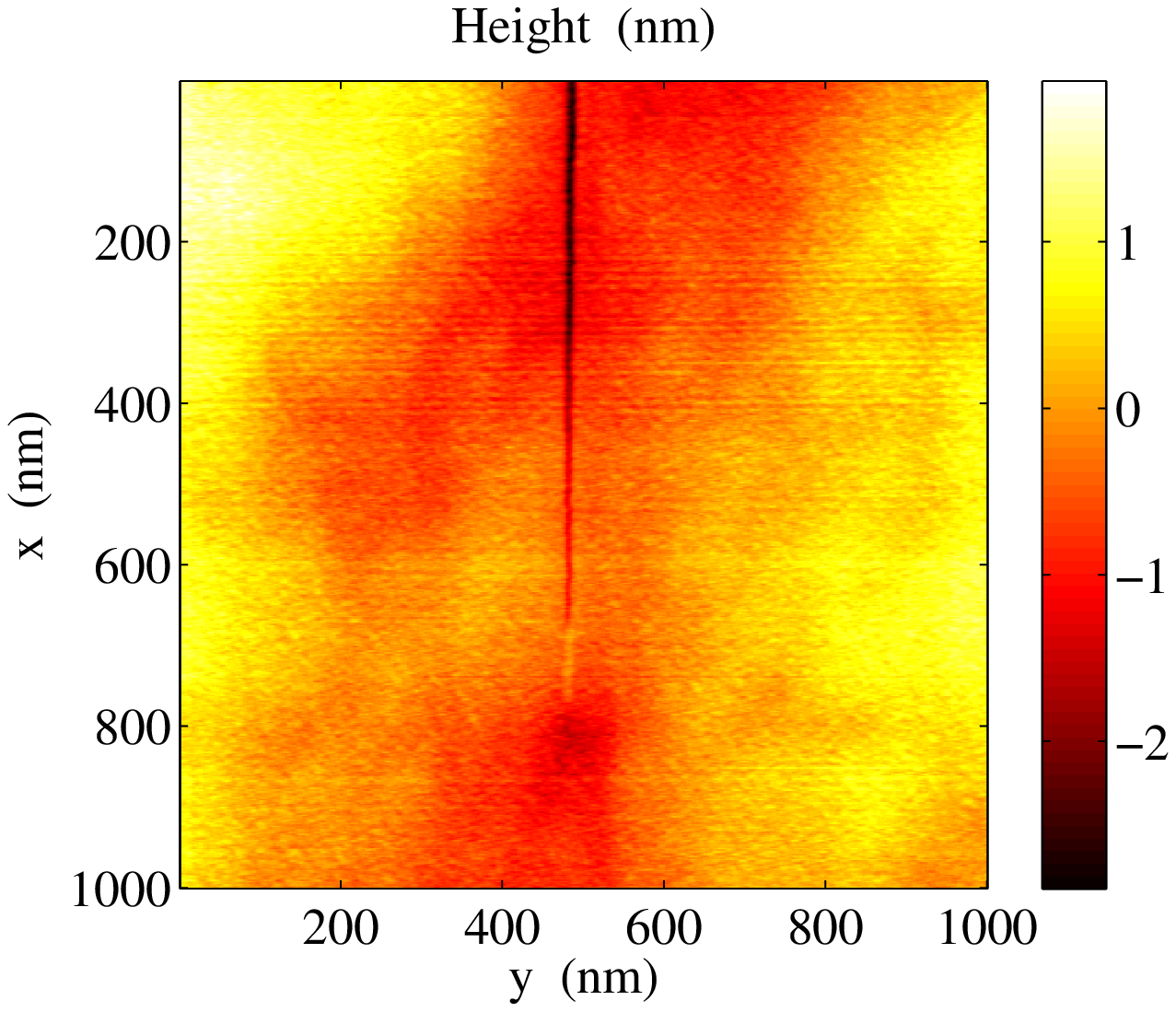}
b)\includegraphics[width=0.45\columnwidth]{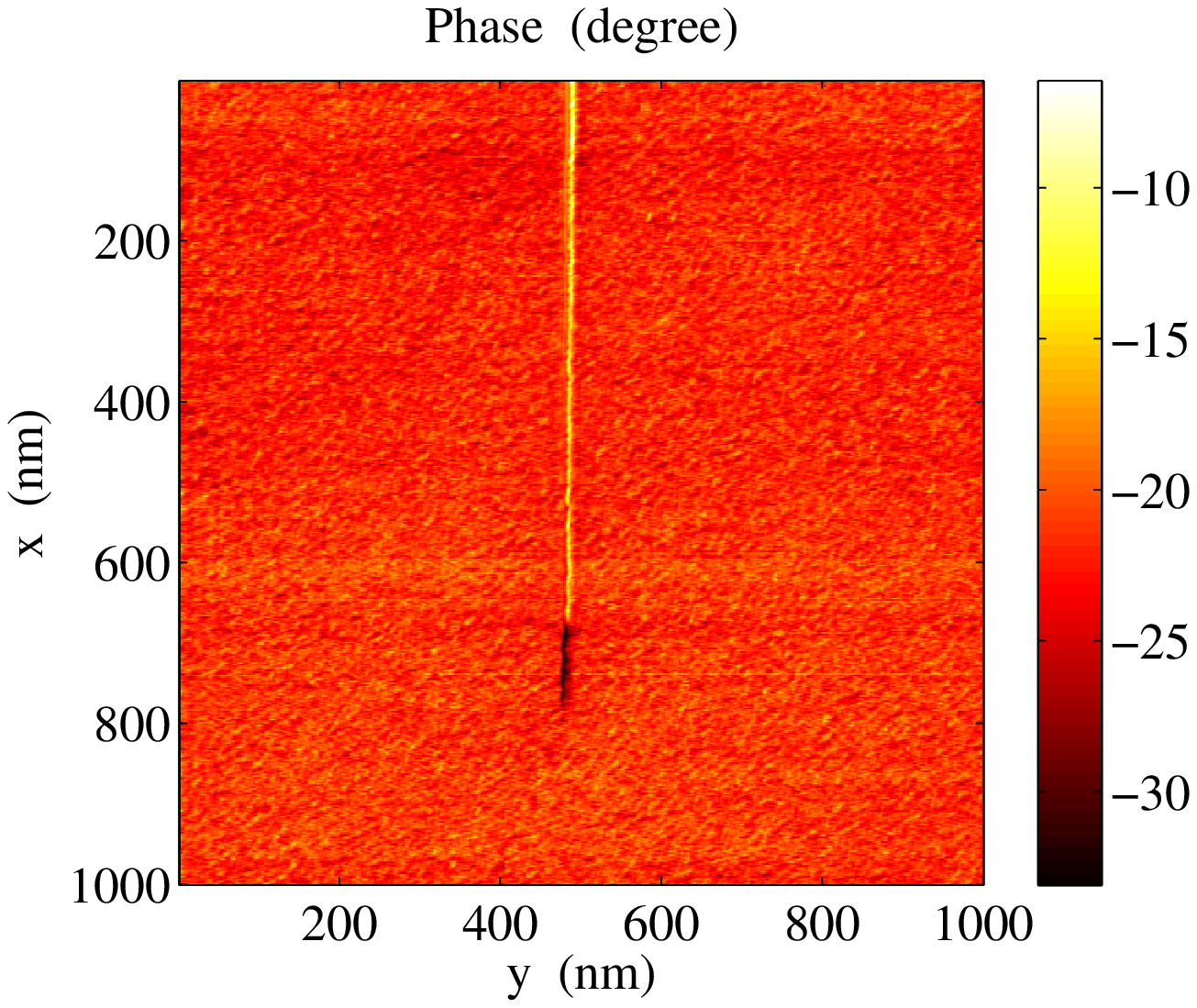}
\caption{\label{Fig:AFMref_topo&phase} Topographic (a) and phase
(b) images.  The whole crack path appears to be clearly visible on the
phase map, but the crack tip can hardly be perceived on the topographic image.}
\end{figure}

Finally, the above computation requires the knowledge of the crack tip location.
If, at a large scale, the crack path is clearly visible on the topographic
images, the exact position of the tip is more uncertain (cf.\ Fig.\
\ref{Fig:AFMref_topo&phase}(a)). However, the hydrophilic nature of the glass
surface induces a capillary condensation of water at the crack tip, and when the
AFM tip probes the immediate vicinity of the crack, the formation (and
disruption) of a water meniscus induces a significant change in the phase angle
measurements \cite{Grimaldi} as can be seen in
Fig.~\ref{Fig:AFMref_topo&phase}(b). Thus, albeit phase angle maps are much more
homogeneous than topographic ones, and hence can hardly be used for IDIC, the
crack path and tip can be resolved precisely on those images. We should note
that the present phase images are acquired using the repulsive tapping mode,
which allows for a resolution of 10 nm on the lateral positioning of the crack
tip, but good results  can also be obtained using either the attractive mode or
a spatially mixed mode between the surface and the condensate (cf.
\cite{Wondrakzec,Grimaldi}). In addition, starting from these observations of
the crack tips on both reference and deformed images, a final optimization can
be performed based on the minimization of the functional $\cal T$.

One major difficulty is that the analytical form of the out-of-plane field
$\psi(\bm x)$ is unknown.  The general problem of the emergence of a crack on a
free face, is a difficult issue. In the framework of linear elasticity, at the
very point of intersection between the crack front and the free surface, \ie the
observed ``crack tip'', a point singularity occurs, whose exponent depends
continuously on the angle of the crack front incidence to the external surface
and on the  Poisson's ratio \cite{Benthem}. A detailed theoretical and numerical
study of this effect applied to glass and specifically in the very same DCDC
test as the one here considered can be found in Ref.~\cite{Fett09}.

It has been argued~\cite{Dimitrov} that the crack front geometry adjust itself
at propagation so that the singularity exponent is the same, $1/2$, as for a
planar crack. Thus, as a rough approximation, we first used $\psi_0(\bm
x)=\sqrt{|\bm x|}$. Although, not equal to the actual displacement field, yet it
allowed to register both images, with the appropriate in-plane displacement.
Once this is achieved, we can extract the height difference $\delta z(\bm x)$
between the reference and deformed surfaces, ignoring the $\psi_0(\bm x)$
contribution. This difference (or ``residual'') should be expressed as $\psi(\bm
x-\bm x_0)-\psi(\bm x-\bm x_1)$. A simple algebraic expression for $\psi$ is
proposed as a systematic Fourier expansion respecting the mode I crack
symmetries
  \be\label{eq:fit_out}
  \psi(\bm x)=\sqrt{|\bm x|}\sum_{n} \alpha_{n}\cos(n\theta/2)
  \ee
where $\theta$ is the polar angle of the current point $\bm x$ with
respect to the crack tip. Low orders for the parameter $0\le n\le 2$
are chosen.  This allows for a direct experimental determination of
the out-of-plane displacement field which is then used in the IDIC
analysis.



\begin{figure}[!ht]
a)\includegraphics[width=0.45\columnwidth]{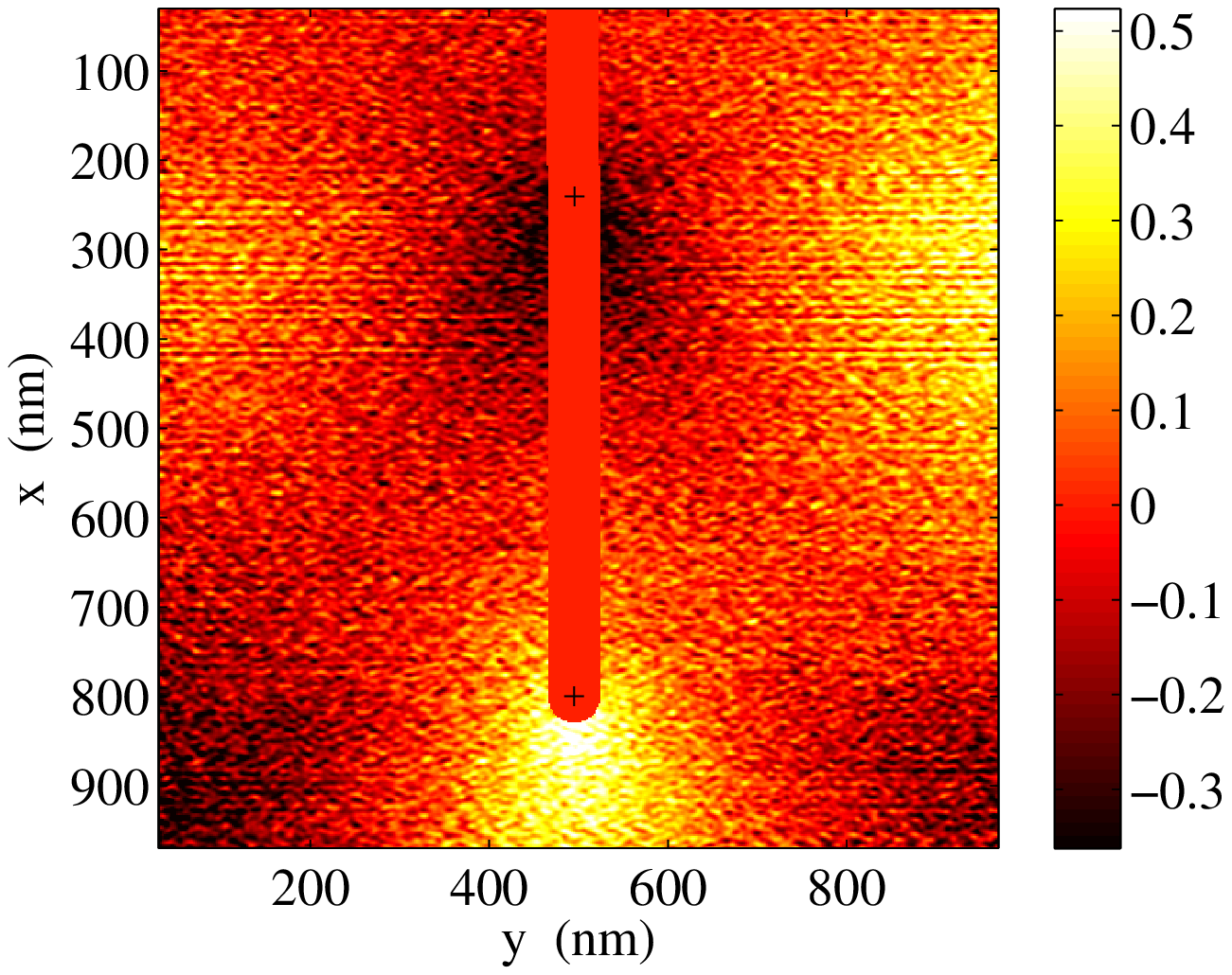}
b)\includegraphics[width=0.45\columnwidth]{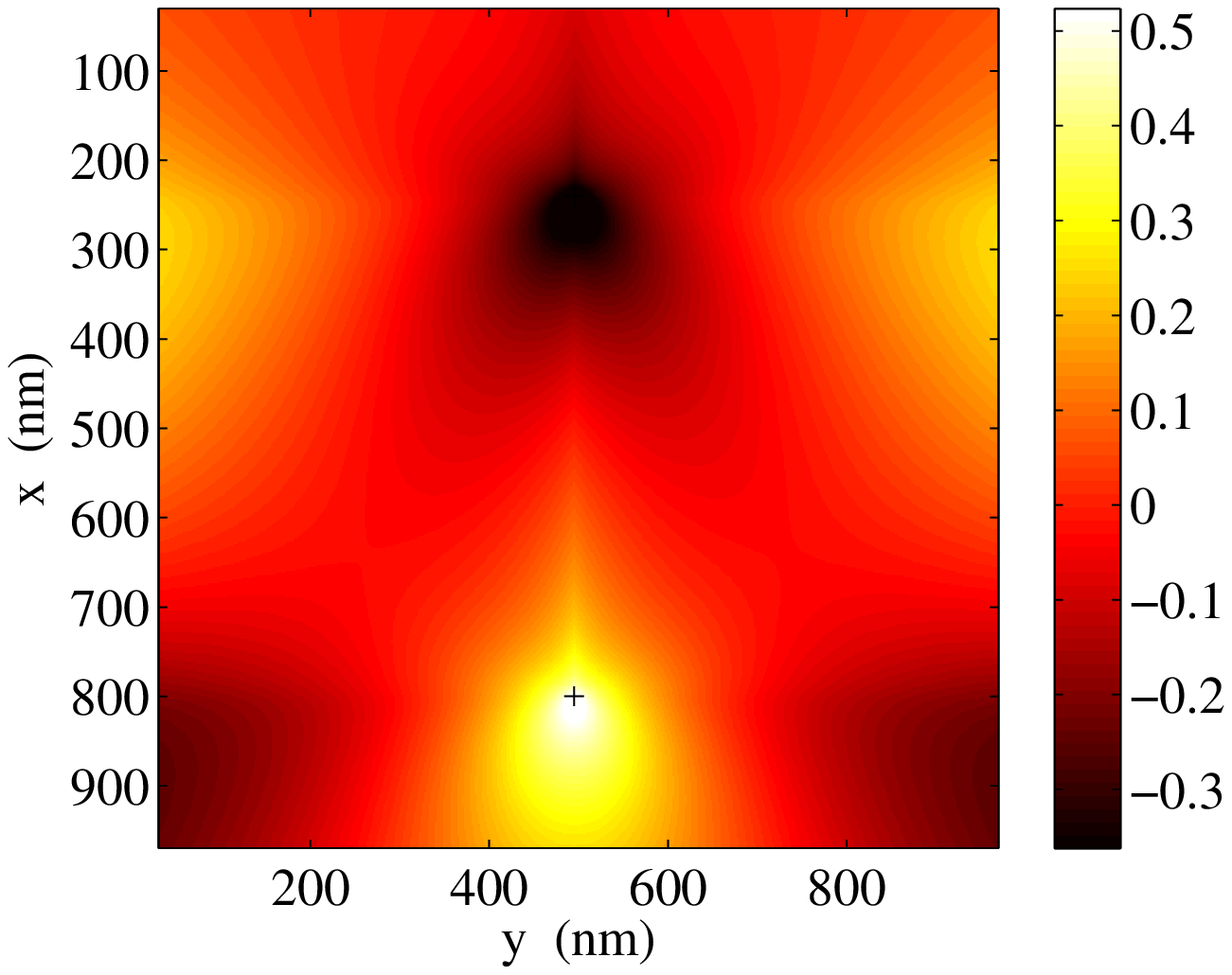}
c)\includegraphics[width=0.45\columnwidth]{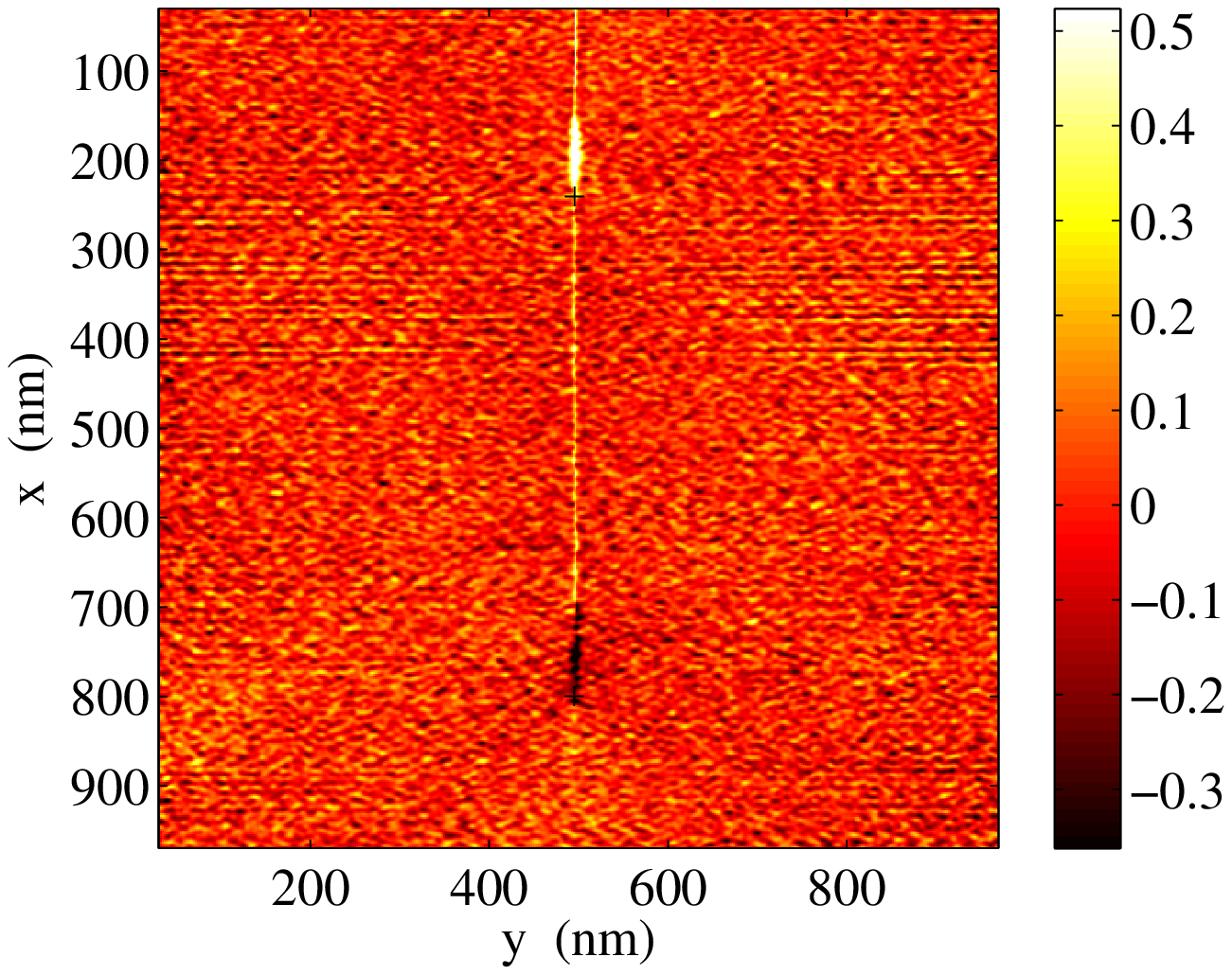}
d)\includegraphics[width=0.45\columnwidth]{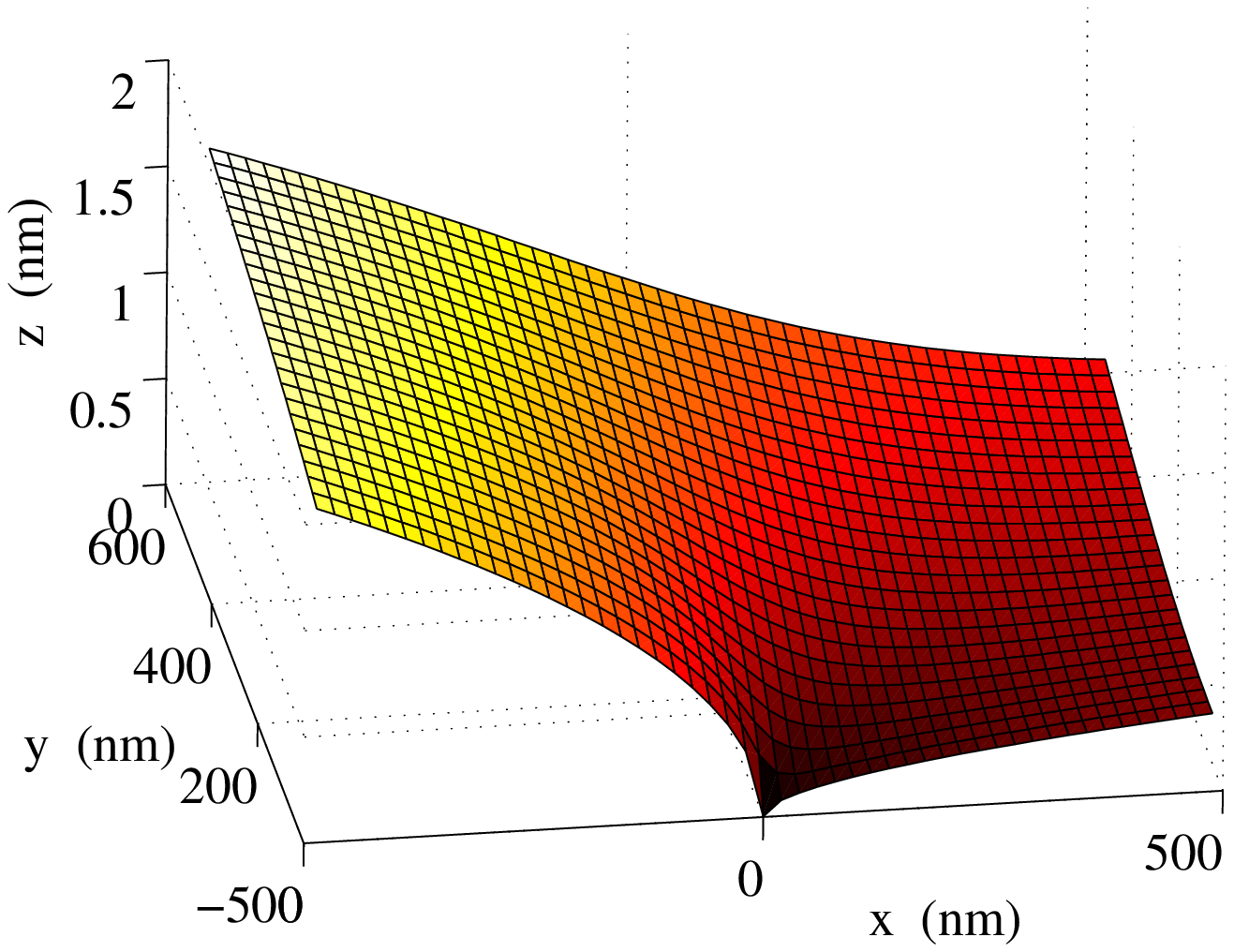}
\caption{\label{Fig:AFMref_Out} (a) Raw residual field
obtained after correction of in-plane displacement. Note that
the crack path is masked for the IDIC analysis; (b)
Fitted displacement field, $\psi(\bm x-\bm x_0)-\psi(\bm x-\bm
x_1)$. (c)
Remaining residual after fit subtraction.  The mask
has not been included to make the crack visible; (d)
Out-of-plane displacement $\psi(\bm x)$ associated with the
crack opening.
A cut through the symmetry plane $y=0$ is
shown, with the crack tip located at the origin.
}
\end{figure}

A series of 5 images of the same $1\times 1$ $\mu$m$^2$ zone swept by
a crack is first analyzed. An example of the residual map without
out-of-plane displacement is shown in Fig.~\ref{Fig:AFMref_Out}(a).
From the set of the four residual fields, a least squares
regression is implemented in order to identify the out-of-plane
displacement based on the algebraic expression Eq.~\ref{eq:fit_out}.
$\alpha= (-0.39,-0.94,1.0)$ is obtained.
Figure~\ref{Fig:AFMref_Out}(c) shows that no obvious long range trend
is left in the final residual. The out-of-plane displacement due to
the opening of a crack at the origin is shown in the same Figure.  It
is to be emphasized that the present analysis cannot determine the
absolute reference plane and thus $\psi$ is determined up to a linear
function in $\bm x$.  Yet, the order of magnitude of the displacement
amplitude is quite comparable to the result of numerical simulations
reported by Fett \etal~\cite{Fett09}.

This expression of the out-of-plane displacement is now used in the sequel. The
quality of the IDIC measurement is evaluated by measuring the normalized
residual, $\eta$, \ie the standard deviation of the remaining residual,
normalized by the peak-to-valley roughness of the original topographic image. In
the present cases, even though the kinematic field was adjusted with very few
parameters, the normalized residual $\eta$ was very stable at about 1.6\%, a
remarkably low value which gives confidence in the convergence of the procedure.
The measured SIF values (assuming a Young's modulus of 72 GPa, and a Poisson's
ratio of 0.17) were successively  $K_I=$ 0.38, 0.36, 0.44 and 0.38
MPa.m$^{1/2}$. The stability of the four independent measurements gives an {\em
a posteriori} indication on the uncertainty of the toughness measurement,
$0.39\pm 0.04$ MPa.m$^{1/2}$, which is in excellent agreement with the
macroscopic value of $0.39\pm 0.02$ MPa.m$^{1/2}$ estimated independently.  Note
the huge difference of length scales used in both estimates (millimeters {\it
vs.} nanometers respectively for the macroscopic evaluation and present
analysis).




\begin{figure}
a)\includegraphics[width=0.45\columnwidth]{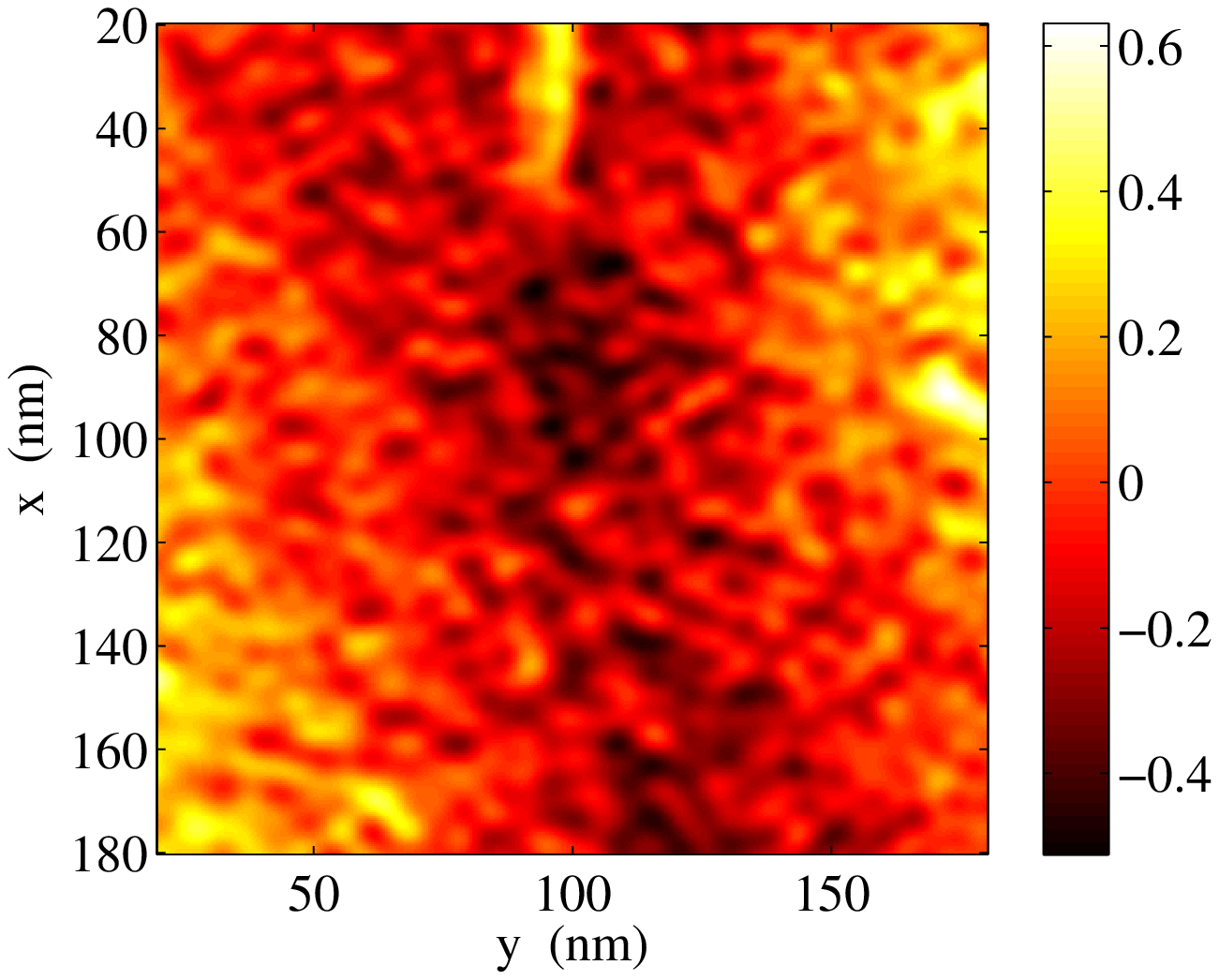}
b)\includegraphics[width=0.45\columnwidth]{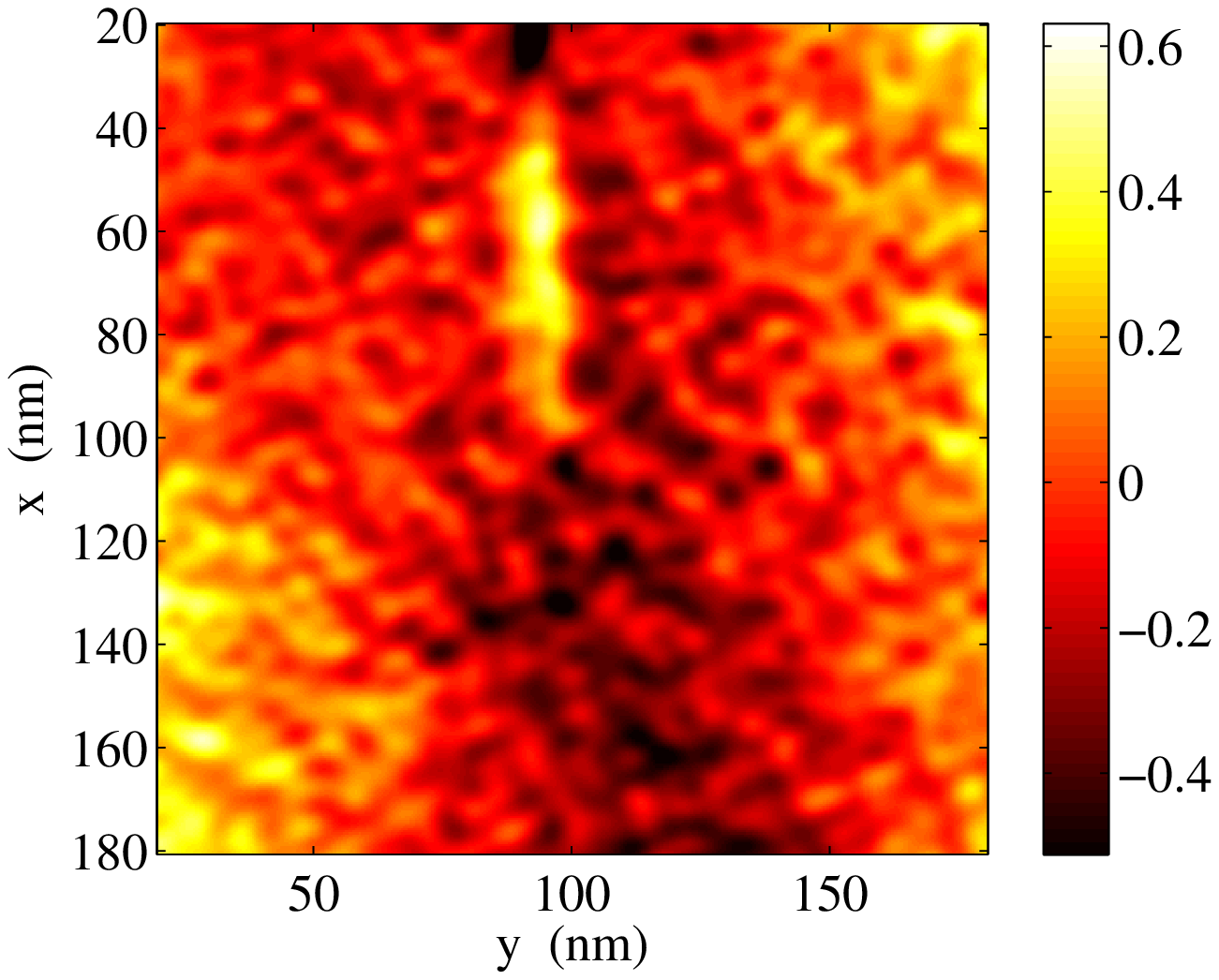}
c)\includegraphics[width=0.45\columnwidth]{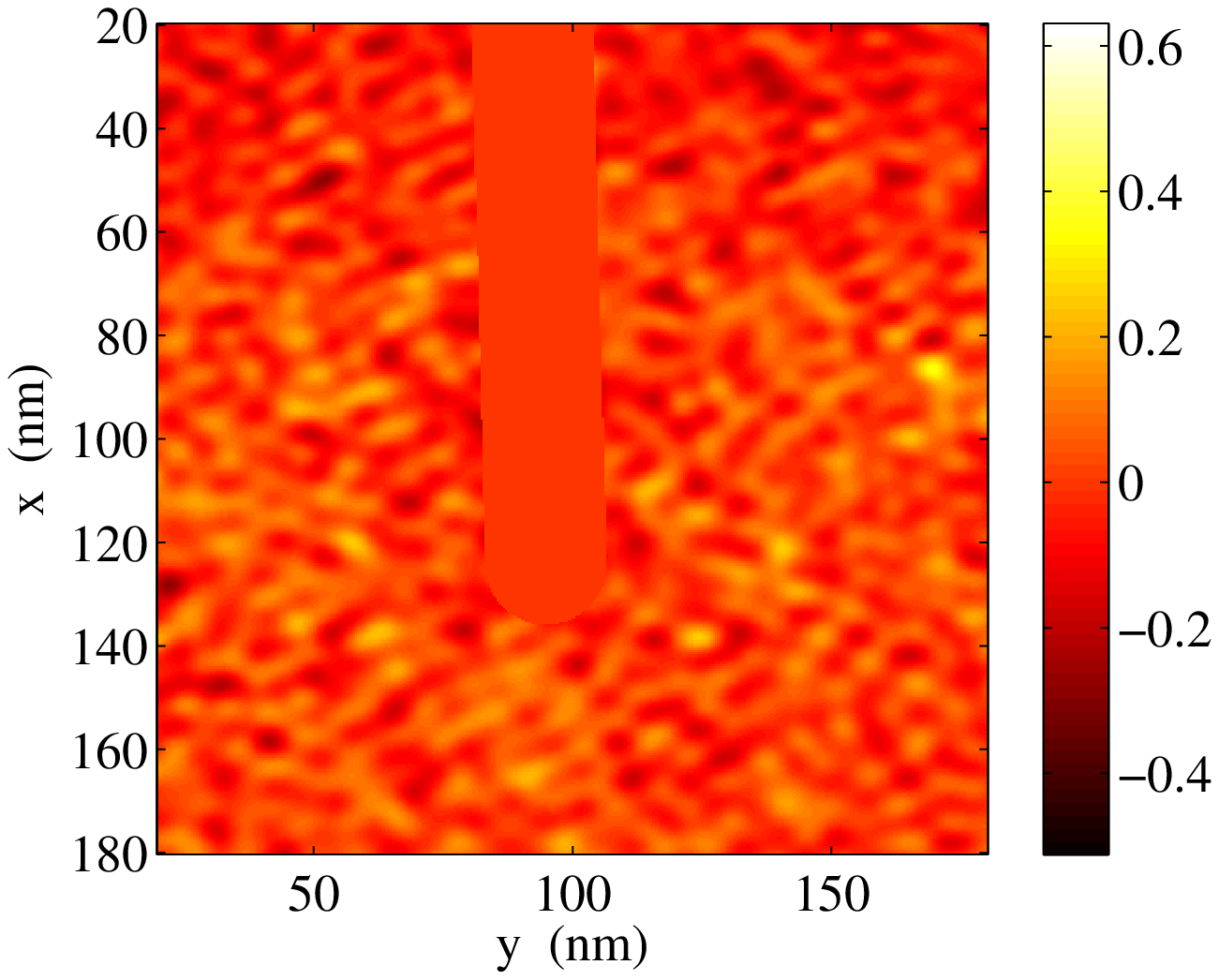}
d)\includegraphics[width=0.45\columnwidth]{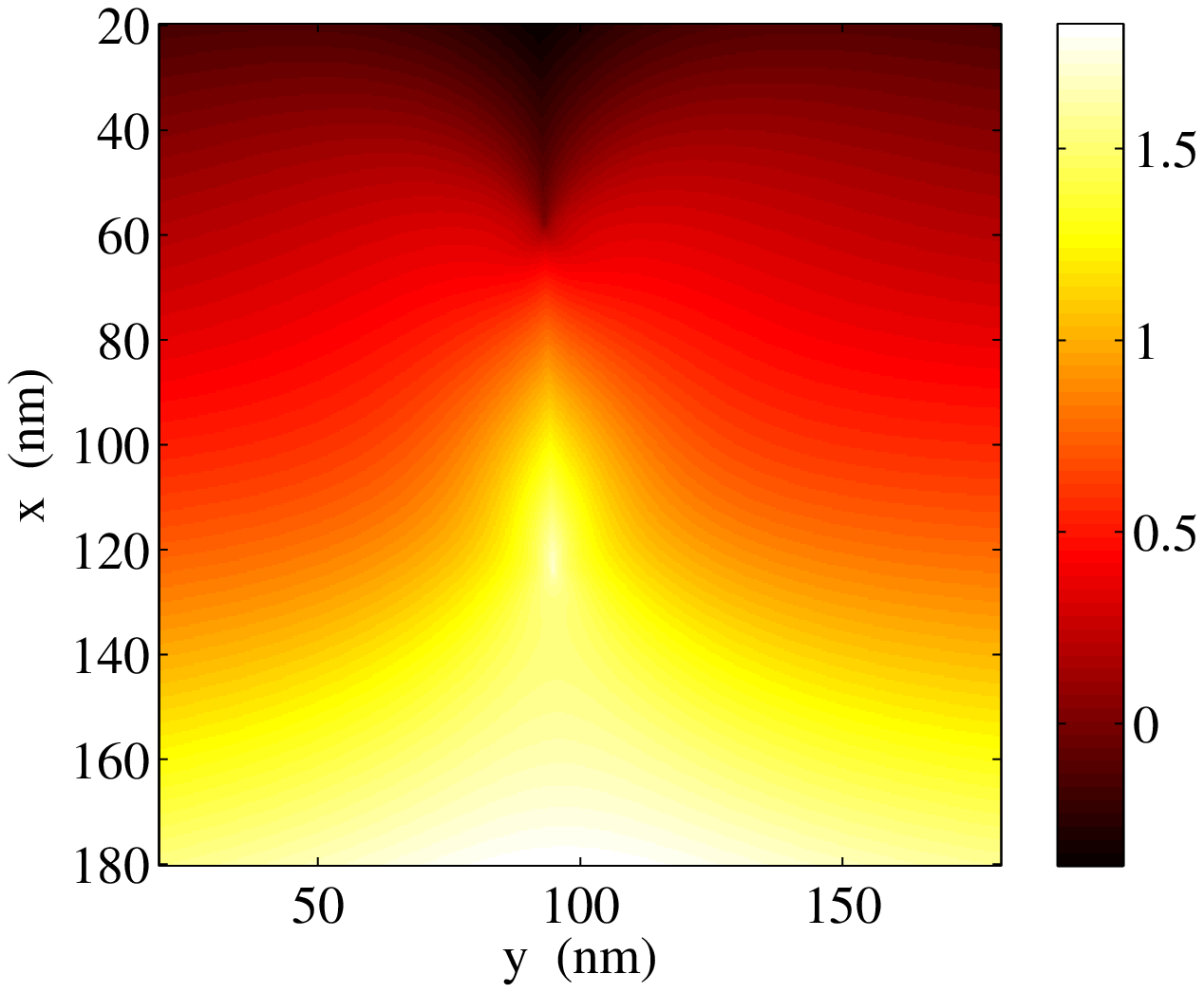}
e)\includegraphics[width=0.45\columnwidth]{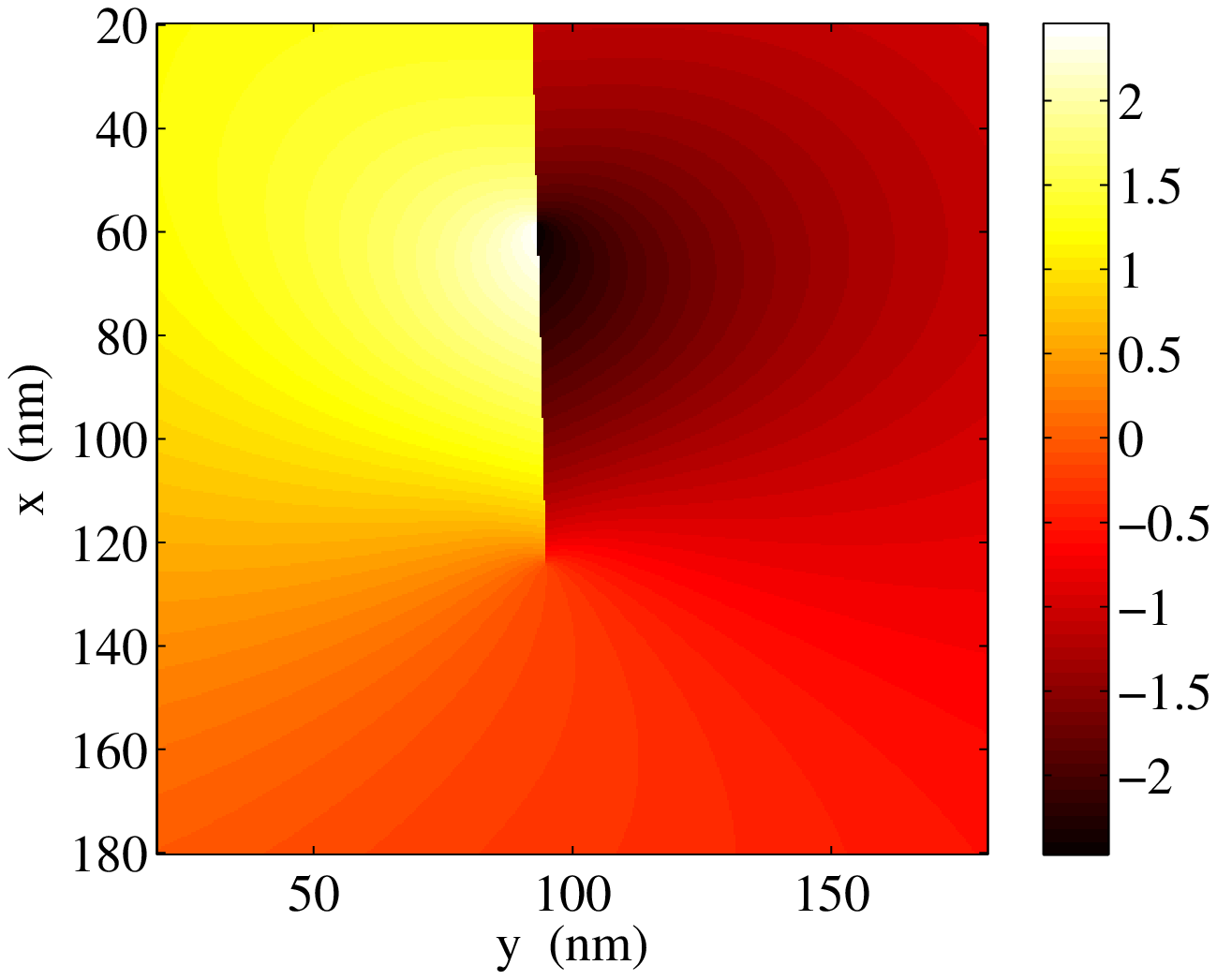}
f)\includegraphics[width=0.45\columnwidth]{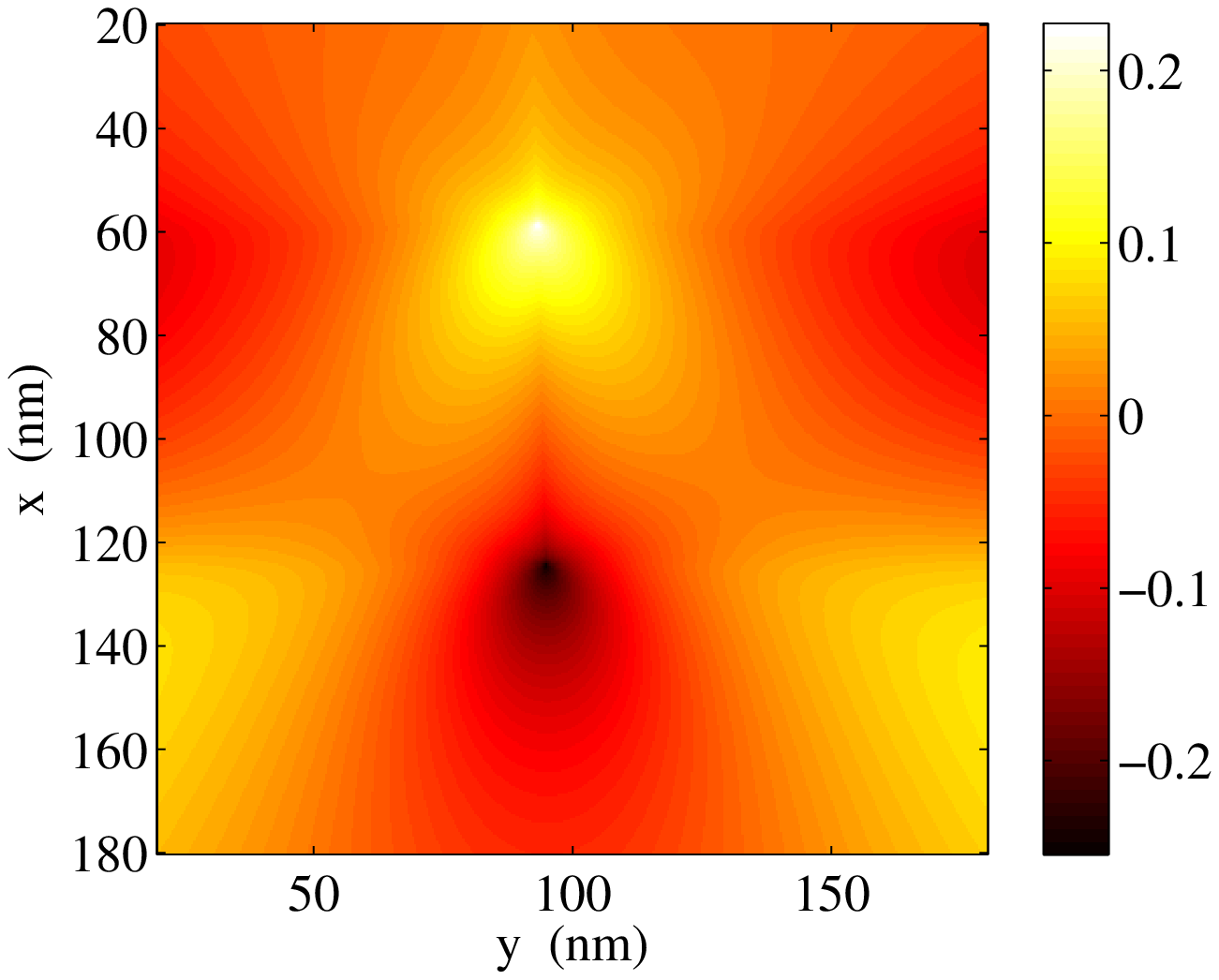}
\caption{\label{Fig:Ana_200nm}  Analysis of 200 nm size
images.
(a): Reference image;
(b): Deformed image;
(c): Residual map (where the masked region around the crack can be seen);
(d): $U_x$ displacement map (in nm);
(e): $U_y$ displacement map (in nm).
(f): $U_z$ displacement map (in nm);
}
\end{figure}

Going to the smaller scan size of 200 nm, a number of additional difficulties
arises: First the precise reposition of the same zone is done within a few tens
of nanometers, i.e. a large fraction of the scanned image and such a large
offset is more difficult to handle for DIC.  This difficulty was taken care of
by a prior rough determination of the mean translation. Second, the fluctuations
in the ``slow scan'' direction are much more intense than in the fast scan
direction (here transverse to the cracks).  This high frequency noise of
amplitude 0.9 {\AA} is only partly corrected by the ``flattening'' filter used and
it makes the computation of image gradient insecure. To further limit this
noise, a low-pass Fourier filtering was applied on the image at the cut-off
frequency for which the slow and fast scan power spectra depart from each other.
For 200 nm images, although being composed of 512$\times$512 pixels, wavelengths
below 4 pixels (\ie about 1.5 nm) were filtered out. Another major difficulty
comes from uneven step sizes in the slow scan direction. A simple bare eye
inspection does reveal that indeed, images appear to be either dilated or
expanded along the slow scan direction by a significant amount.  Similarly, due
to inaccuracies in the reposition at the beginning of each new line an
artificial simple shear can be introduced.  Thus, we implemented additional
degrees of freedom in the kinematics, (i.e. uniform and linearly varying strain
along the slow scan direction, and uniform shear strain) corresponding to these
artefact deformation modes. Finally, it is worth mentioning that pointing at the
crack tips, even from phase images, is quite delicate and does contribute to the
final uncertainty.

With those additional procedures, image series could be analyzed
successfully, in the sense that the residual, $\eta$, could be
brought down to 5-6\%.  A visual check based on the comparison
between reference and corrected deformed images indeed confirms the
reliability of the image matching, and hence the robustness of the
algorithm, in spite of the significant level of noise in the images.
Measurements of the mean strain along the slow-scan direction
revealed strains which could be as large as 8\%. In the example shown
in Fig.~\ref{Fig:Ana_200nm}, the SIF is estimated to be 0.44
MPa.m$^{1/2}$, the axis strain is 1.6\%, and the residual amounts to
$\eta=5.4\%$.  Ignoring the most strained image pair, the SIF value
was estimated in the range $0.37<K_I<0.46$ MPa.m$^{1/2}$. Varying the
size of analyzed domain, and the cut-off frequency induces only very
slight modifications in the estimates of the SIF.  The estimate
$K_I=0.41\pm 0.05$ MPa.m$^{1/2}$ is again consistent with the
macroscopic determination and the larger scale AFM IDIC estimate.
Most of the uncertainty lies in the precise identification of the
crack tip location.
As can be seen in the residual map shown in
Fig.~\ref{Fig:Ana_200nm}(c), no systematic bias can be
distinguished even in the vicinity of the crack tips.

Based on those quantitative observations, no evidence of a breakdown of linear
fracture mechanics --- eventually due to micro-damage or plasticity
--- could be observed even in the immediate vicinity (within 10 nm)
of the crack tip. The present IDIC technique has thus proven to be suited
to measuring stress-intensity factors in brittle materials
by comparing pairs of AFM images of the external sample surface
taken at different stages of crack propagation.
Moreover, this analysis gives an original insight into
the out-of-plane displacement fields of the free surface
in the neighborhood of a propagating crack tip,
which is presently an open subject of research.

\acknowledgments
It is a pleasure to acknowledge useful discussionss
with F. Hild, C. Marli\`{e}re, F. C\'elari\'e, T. Fett and S. M. Wiederhorn.
This work was supported by the ANR project CORCOSIL (BLAN07-3\_196000).

\end{document}